\documentclass[notitlepage,pra,showpacs,amsmath,amssymb,longbibliography]{revtex4-1}

\usepackage{graphicx}
\usepackage{latexsym}
\usepackage{graphicx}
\usepackage{times}
\usepackage{color}
\usepackage{amsmath}
\usepackage{dcolumn}
\usepackage{mathbbol}
\usepackage[final]{pdfpages}
\usepackage{latexsym,amsmath,amssymb,bm,euscript}

\newcommand{\ig}[1]{}

\newcommand{\changes}[1]{{\color{black}{#1}}}

\newcommand{\da}{\dagger}
\newcommand{\ki}{ \alpha \mathbf{k}}

\newcommand{\kq}{\mathbf{k}\mathbf{q}}

\newcommand{\kr}{\rm{L}\mathbf{q}}
\newcommand{\kl}{\rm{U}\mathbf{k}}

\begin{document} 

\title{Cavity squeezing by a quantum conductor} 

\author{Udson C. Mendes}
\author{Christophe Mora}
\affiliation{Laboratoire Pierre Aigrain, \'Ecole Normale Sup\'erieure-PSL Research University, CNRS, Universit\'e Pierre et Marie Curie-Sorbonne Universit\'es, Universit\'e Paris Diderot-Sorbonne Paris Cit\'e, 24 rue Lhomond, 75231 Paris Cedex 05, France}
\email{mora@lpa.ens.fr}

\date{\today} 

\begin{abstract} 
Hybrid architectures integrating mesoscopic electronic conductors with resonant microwave cavities have a great potential for investigating unexplored regimes of electron-photon coupling. In this context, producing nonclassical squeezed light is a key step towards quantum communication with scalable solid-state devices. Here we show that parametric driving of the electronic conductor induces a squeezed steady state in the cavity. We find that squeezing properties of the cavity are essentially determined by the electronic noise correlators of the quantum conductor. In the case of a tunnel junction, we predict that squeezing is optimized by applying a time-periodic series of quantized $\delta-$peaks in the bias voltage. For an asymmetric quantum dot, we show that a sharp Leviton pulse is able to achieve perfect cavity squeezing.

\end{abstract} 


\maketitle 

\section{Introduction}
Squeezed states of light~\cite{walls1983} exhibit reduced noise below the vacuum level in one of their quadrature and amplification in the other quadrature. Their realization is a key step in the development of quantum communication. They are important tools for continuous variable quantum information protocols~\cite{braunstein2005,menzel2012} where they serve as building blocks for generating non-classical states. Their enhanced sensitivity can also be used for quantum non-demolition measurements of position and force~\cite{caves1980}. Easily produced in optical systems, squeezed states have been observed more recently in circuit quantum electrodynamics at microwave frequencies~\cite{yurke1988}, either as single-mode~\cite{castellanos2008,mallet2011}, two-mode squeezing~\cite{eichler2011,wilson2011} or as Einstein-Podolsky-Rosen states~\cite{flurin2012}.

The parametric driving used so far in experiments is limited to a half-squeezed quadrature for a cavity mode because of the inevitable coupling to the external vacuum fluctuations~\cite{milburn1981}. This limit however does not apply to dissipative squeezing, in which one steers the environment to stabilize the cavity into a non-classical state. In this case, perfect squeezing can be achieved, at least in principle, with minimum uncertainty~\cite{kronwald2013,didier2014}.

A recent development in the field of superconducting quantum circuits is the realization of hybrid systems in which a quantum conductor is coupled to a microwave resonator. These systems offer an appealing platform for investigating fundamental matter-light interactions with an experimental control on both the electronic and photonic parts~\cite{cottet2011,BergenfeldtSET2012,Souquet2014,schiro2014,cottet2015}. Experiments have been realized with metallic tunnel junctions connected to a resonating line~\cite{altimiras2014}, or with quantum dots, realized in carbon nanotubes\ig{~\cite{delbecq2011,delbecq2013,Viennot2014}}, nanowires\ig{~\cite{Petersson2012,Liu2014}} or two-dimensional electron gases\ig{~\cite{frey2012,toida2013,basset2013}}~\cite{delbecq2011,delbecq2013,Petersson2012,frey2012,basset2013,Viennot2014,Liu2014} embedded in high-finesse coplanar cavities. The interplay of electron transport and emission of photons can lead to an electronic-induced lasing state in the cavity~\cite{Lasing2011,xu2013,DT72014,liu2015}, and more generally produce bunched or antibunched photons~\cite{beenakker2001,beenakker2004,gabelli2004,lebedev2010,Jin2015}, and nonclassicality in the light emitted by a quantum conductor~\cite{CottetMajos2013,bednorz2013}. Squeezed light emitted by a tunnel junction was recently demonstrated experimentally in the absence of a cavity~\cite{gasse2013,forgues2014,forgues2015}.

\begin{figure}
\begin{center}
\includegraphics[width=0.5\textwidth]{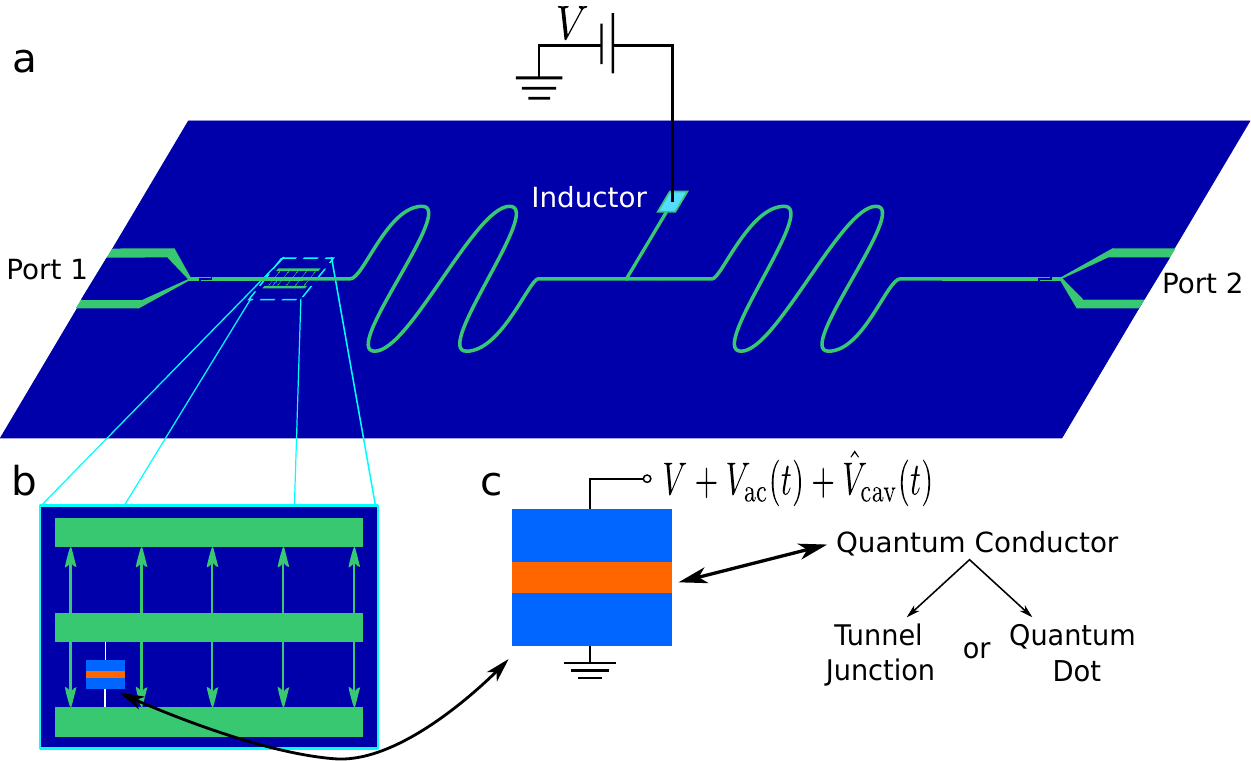}
\caption{(a) Schematic of a resonant cavity realized in a superconducting transmission line. The center line is coupled to external electromagnetic modes (ports) via capacitive gaps delimiting the cavity. The output field of the cavity can be measured in these ports. A dc voltage can be applied to the center line at a voltage node in order to preserve the high-Q factor of the cavity~\cite{Petersson2012,chen2014}. (b) Focus on the galvanic coupling between the cavity and a quantum conductor. The electric field (arrows) is spatially constant at the scale of the conductor. The conductor is lead-contacted on the upper side to the center line and on the lower side to the ground plane. (c) The quantum conductor can be a tunnel junction or a quantum dot. The upper lead is connected to the center line where the voltage potential is the sum of  dc [$V$] and ac [$V_{\rm ac}$(t)] parts, as well as the cavity quantum field $\propto a+a^\dagger$.  \label{fig1}}
\end{center}
\end{figure}

In this paper, we describe dissipative squeezing of a cavity mode coupled to an ac driven electronic reservoir. The system, depicted in Fig.~\ref{fig1}, is a quantum conductor coupled to a microwave resonator. In addition, a classical bias voltage, with an oscillating part at twice the resonator frequency, is applied to the conductor. The conductor plays the role of a nonlinear environment: photons from the ac modulation are broken into pairs and transmitted to the cavity, thereby producing squeezing. We show that the amount of cavity squeezing is determined solely by current noise fluctuations in the conductor. We focus our attention on ac excitations which optimizes squeezing. In the case of a tunnel junction, we find that the best solution consists of periodic and quantized voltage peaks occurring in phase with the compressed quadrature. We also discuss how squeezing improves with the number of harmonics in the ac signal. For a quantum dot, we identify the conditions of optimum squeezing: asymmetric coupling to the leads, a narrow single-level resonance, a far-detuned single-level energy and a dc bias voltage matching the resonator frequency. In addition, we show that a Leviton pulse results in a vacuum squeezed state with minimal uncertainty. Perfect squeezing is approached by narrowing the width of the voltage pulses in the Leviton, again in phase with the compressed quadrature.

\section{Tunnel junction}
 We start by considering the case of a tunnel junction for the quantum conductor. The voltage bias across the junction is the sum of a driven classical part applied on the upper lead, and a quantum part associated to the cavity field. Using a convenient choice of electromagnetic gauge, it simply dresses the electron tunneling operators
\begin{equation}\label{eq-tunn}
H_{\rm T} (t) = {\cal T} e^{-i \hat{\varphi} (t)} + {\cal T}^\dagger e^{i \hat{\varphi} (t)},
\end{equation}
where the operator ${\cal T}$ transfers one electron from the upper to the lower lead while ${\cal T}^\dagger$ does the opposite~\cite{safi2010,safi2015}. $ \hat{\varphi} (t)$ is the total semi-classical phase accumulated during a tunneling event, decomposed as $ \hat{\varphi} (t) =  (e V/\hbar) t +\phi_{\rm ac} (t) + i (g/\omega_0) (\hat{a}^\dagger - \hat{a})$, with the ac phase $\phi_{\rm ac} (t)=  (e/\hbar) \int^t d t' \, V_{\rm ac}  (t')$. The last term is the quantum part and $g$ measures the junction-cavity coupling strength. The cavity Hamiltonian is reduced for simplicity to a single mode, $H_{\rm cav} = \hbar \omega_0 \hat{a}^\dagger \hat{a}$ with the cavity annihilation operator $\hat{a}$. Eq.~\eqref{eq-tunn} contains both the excitation of the cavity state by electron tunneling events and photo-assisted transport phenomena triggered by the ac modulation~\cite{tien1963,Kouwenhoven1994}.  $H_{\rm T}(t)$ is thus responsible for an exchange of energy between three sub-systems: the cavity, the ac classical field and lead (free) electrons.

We assume weak junction-cavity coupling and therefore neglect the backaction-induced change in electron tunneling resulting from the cavity. We thus set $g=0$ to examine the current fluctuations of the tunnel junction and latter reinstate a finite $g$ when considering the dynamics of the cavity field $\hat{a}$. In the presence of ac voltage modulation, the photo-assisted noise properties of the tunnel junction are characterized by the correlator (at $g=0$)~\cite{gabelli2008}
\begin{equation}\label{noise}
\langle \hat{I} (\omega_1) \hat{I} (\omega_2) \rangle = \sum_{n=-\infty}^{+\infty} S_n (\omega_1) 2 \pi \delta(\omega_1+\omega_2 - 2 n \omega_0),
\end{equation}
where $\hat{I} (\omega)$ is the Fourier transform of the Heisenberg current operator $\hat{I} (t)$ of the junction. The $n=0$ term gives the stationary part of the noise, meaning that, when Fourier transformed with times $t_1$ and $t_2$, it depends only on the time difference $t_1 - t_2$ and not on the mean time $\bar{t} = (t_1+t_2)/2$.  $S_0 (\omega)$ is called the absorption (emission) noise of the tunnel junction for $\omega>0$ ($\omega<0$)~\cite{basset2010}. It governs the rate of energy transfer between the junction and its environment, here the cavity, via single photons of energy $\hbar |\omega|$. 

As shown in the Supplementary Note 1, the different noise terms can be calculated to leading order in the tunneling strength
\changes{
\begin{equation}\label{noise-terms}
2 S_n (\omega) = \sum_{n'} \left[ c_{n'} c_{n'+n}^* \bar{S} (\omega + e V/\hbar + 2 n' \omega_0) 
 +  c_{n'}^* c_{n'-n} \bar{S} (\omega - e V/\hbar - 2 n' \omega_0) \right] e^{-i n \varphi},
\end{equation}}
in terms of the (unsymmetrized) equilibrium Johnson-Nyquist noise of the tunnel junction $\bar{S} (\omega) =  (2 \omega /R_T) (1-e^{-\hbar \omega/(k_B T)})^{-1}$, where $R_T$ is the junction dc resistance. \changes{The coefficients $c_m$ are defined by the Fourier expansion
\begin{equation}\label{coefffourier}
e^{i \phi_{\rm ac} (t)} = \sum_{m \in \mathbb{Z}} c_m e^{2 i m  \omega_0 t} e^{i m \varphi},
\end{equation}
of the ac phase $\phi_{\rm ac}$. $\varphi$ is the overall phase of the ac signal.  For the lead connected to the voltage $V_{\rm ac}(t)  = (\hbar/e) \dot{\phi}_{\rm ac}(t)$, $c_m$ gives the probability amplitude for an electron to absorb $m$ energy quanta from the classical ac field when $m>0$. $m<0$ describes correspondingly photon emission to the ac field~\cite{dubois2013}. In the particular case of a sinusoidal excitation, $V_{\rm ac}  (t) = V_1 \cos( 2 \omega_0 t + \varphi)$, these coefficients are written in terms of Bessel functions 
\begin{equation}
c_m = J_m \left( \frac{e V_1}{2 \hbar \omega_0} \right).
\end{equation}}
The energy of absorbed photons, $2 m \hbar \omega_0$, can be either used in exciting energetic electron-hole pairs in the conductor, or transferred to the cavity. The non-stationary noise terms $S_{n \ne 0}$ in Eq.~\eqref{noise} oscillate with the mean time $\bar{t}$. They do not conserve energy and can provide $n$ quanta of energy $2 \hbar \omega_0$ (or absorb if $n<0$) to the cavity. We will return below to the physical significance of these terms when analysing the cavity stationary state and squeezing effects.


Now that we have detailed the possible transfers of energy between the ac-excited tunnel junction and its environment (the cavity), we study the cavity evolution under the dissipative influence of the electrons. Assuming weak junction-cavity coupling, we expand $H_T$ to first order in $g/\omega_0$, $H_T = H_T^0 + i \lambda \hat{I} (\hat{a}^\dagger - \hat{a})$ with the coupling constant $\lambda = \hbar g/(e \omega_0)$. The cavity evolution is described by a Heisenberg-Langevin equation
\begin{equation}\label{langevin}
\dot{\hat{a}} + i \omega_0 \hat{a} + \frac{ \kappa}{2} \hat{a} = \lambda  \hat{I} (t),
\end{equation}
\changes{justified, either by an input-output calculation~\cite{other2015} detailed in the Supplementary Note 2, or by a Keldysh path integral formulation, discussed in appendix~\ref{keldysh}, assessing the Gaussian character of current and cavity field fluctuations.} The tunnel junction current $\hat{I} (t)$ \changes{in Eq.~\eqref{langevin}} plays the role of a quantum noise term, with  fluctuations characterized by the correlator of Eq.~\eqref{noise}. In deriving this equation, we have neglected the intrinsic (bare) damping of the cavity $\kappa_0$, assuming that the cavity dissipation caused by the electrons dominates. The corresponding damping rate $\kappa = \lambda^2 [ S_0 (\omega_0) - S_0 (-\omega_0) ]$ balances absorption and emission noises, since absorption (emission) noise corresponds to photon loss (gain) from the cavity. Our calculation is also based on the use of the rotating-wave approximation where rapidly oscillating terms are averaged to zero. The validity of this approximation is controlled by the smallness of $\kappa/\omega_0$ and is consistent with our first order conductor-cavity coupling and with the absence of cavity backaction.

The first order differential Eq.~\eqref{langevin} can  be solved straightforwardly in time or frequency space, and yields the steady-state correlation functions for the cavity field $\hat{a}$. We find for the anomalous correlator, using that $\kappa \ll \omega_0$,
\begin{equation}
\langle \hat{a}^2 \rangle = \frac{\lambda^2S_1 (\omega_0)}{\kappa}=  \frac{S_1 (\omega_0)}{S_0 (\omega_0) - S_0 (-\omega_0)}.
\end{equation}
This result can be given a physical interpretation: $S_1 (\omega_0)$ describes the coherent emission by the junction of a quantum of energy $2 \hbar \omega_0$ to the cavity. This energy quantum breaks into a pair of cavity photons thereby contributing to the $\langle \hat{a}^2 \rangle$ correlator. This effect is limited in the denominator by the rate at which cavity photons are absorbed by the tunnel junction. In the same way, the number of cavity photons $\langle \hat{a}^\dagger \hat{a} \rangle = \lambda^2 S_0 (-\omega_0)/\kappa$ is unsurprisingly governed by the electronic emission noise.

We now investigate field squeezing more precisely and introduce the two quadratures  $\hat{X}_1  = i ( \hat{a}^\dagger e^{-i \varphi/2} - \hat{a} e^{i \varphi/2})$ and $\hat{X}_2 = \hat{a}^\dagger e^{-i \varphi/2} +\hat{a} e^{i \varphi/2}$, where we use the same phase $\varphi$ as in Eq.~\eqref{coefffourier}. Their variance is readily obtained
\begin{equation}\label{quadra}
\Delta X_{1/2}^2 = \frac{\sum_n | c_n \mp c_{n+1} |^2 \bar{S}_e (e V/\hbar + (2 n+1) \omega_0)}{\sum_n \left( | c_n |^2 - |c_{n+1} |^2 \right) \bar{S}_o (e V/\hbar + (2 n+1) \omega_0)},
\end{equation}
where we introduced the even $\bar{S}_e $ and odd $ \bar{S}_o$ parts of the Johnson-Nyquist noise $\bar{S}$. The two quadrature fluctuations are thus sensitive to the electronic temperature $T$, the dc bias voltage $V$ and the pulse shape of the ac signal. The squeezing mechanism is optimized by taking the limit of vanishing temperature and by setting $e V = \hbar \omega_0$. 

We first consider a single-tone driving of the tunnel junction, $V_{\rm ac}  (t) = V_1 \cos( 2 \omega_0 t + \varphi)$, the experimentally most accessible situation. The photo-assisted coefficients $c_n$ are then given by Bessel functions, as detailed in appendix~\ref{appen-leviton}. A numerical minimization of $\Delta X_1^2$ in Eq.~\eqref{quadra} gives $\Delta X_1^2 = 0.618$ for $e V_1 = 0.706 \times 2 \hbar \omega_0$, with $\Delta X_2^2 = 1.864$. This optimal squeezing value coincides exactly with the squeezing of the emitted light predicted and measured in Refs.~\cite{bednorz2013,gasse2013}, \changes{see also the more recent Ref.~\cite{Qassemi2015}}. This is explained by noting that the zero-temperature cavity damping $\kappa/\lambda^2 = 2 \hbar \omega_0/R_T$ (the denominator in Eq.~\eqref{quadra}) is constant for a tunnel junction, regardless of the bias voltage shape. This independence no longer holds at finite temperature or in the case of a conductor with a non-linear {\it I-V} characteristic.

\begin{figure}
\begin{center}
\includegraphics[width=0.5\textwidth]{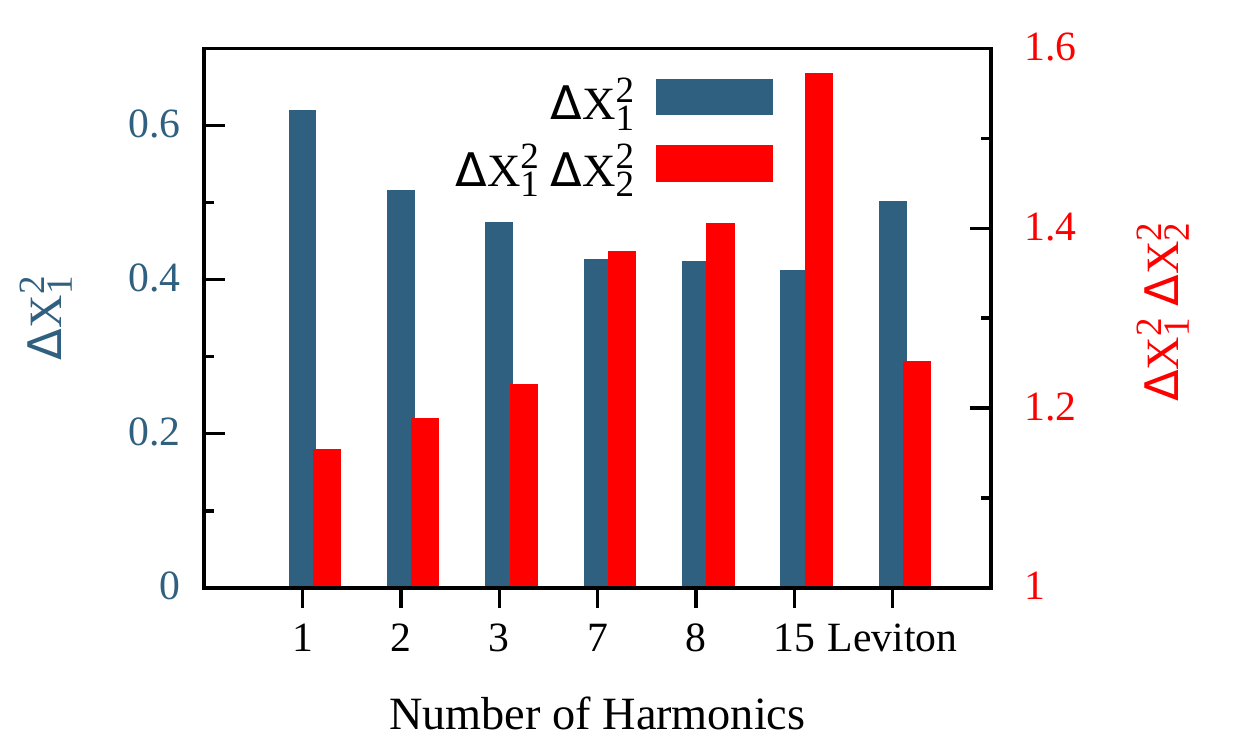}
\caption{Minimum variance of the squeezed quadrature (left) and Heisenberg uncertainty product (right) as a function of the number of harmonics in the ac signal applied to a tunnel junction. The variance $\Delta X_1^2$ of Eq.~\eqref{quadra} is minimized numerically for the ac signal $V_{\rm ac} (t) = \sum_{j=1}^n V_j \cos (2 j \omega_0 t)$ where $n$ is the number of harmonics. Optimal squeezing is obtained with $T=0$, $e V = \hbar \omega_0$. For comparison, the variance has been computed for a Leviton pulse with the result $\Delta X_1^2 = 1 - 2 a+2 a^2$. The optimized Leviton with $a=1/2$ is shown in the last column. \label{fig-sque}} 
\end{center}
\end{figure}

We turn to an ac modulation with the same fundamental frequency $2 \omega_0$ but including higher harmonics~\cite{gabelli2013}. Fig.~\ref{fig-sque} shows the improvement in squeezing $\Delta X_1^2$ by adding more and more harmonics while $\Delta X_2^2$ is further amplified. Considering a general periodic signal, we find analytically, as shown in appendix~\ref{appen-minimum},  that the minimum value $\Delta X_1^2=4/\pi^2 = 0.405$ is reached when $c_n = (1/\pi)(n+1/2)^{-1}$ for $n \in \mathbb{Z}$, in agreement with a numerical minimization. The corresponding ac phase across the junction is a periodic piecewise linear function $\phi_{\rm ac}  (t) = \pi/2-\omega_0 t$ for $t \in ]0,\pi/\omega_0[$, with a jump discontinuity of $\pi$ at $t=0$ and multiples of $\pi/\omega_0$. Adding the dc voltage V, we find for the  optimal voltage applied to the junction a series of $\delta$-peaks, 
\begin{equation}\label{pulse}
V_{\rm opt} (t) = \frac{h}{2 e} \sum_{l \in \mathbb{Z}} \delta\left(t- \frac{l \pi}{\omega_0}\right).
\end{equation}
 It is useful to give an intuitive classical picture for this squeezing optimization: the bias potential $V_{\rm opt} (t)$ acts on the conductor specifically at times where the amplitude of the squeezed quadrature is maximum and the other quadrature vanishes. This is in fact a strong perturbation, the emission and absorption noises are infinite and the second variance $\Delta X_2^2=+\infty$.

Despite its linear current response, the tunnel junction is able to squeeze the cavity state down to $40 \%$ of the zero-point level. This is because squeezing is not governed by the current itself but by current fluctuations, and the noise of a tunnel junction is only a piecewise linear function enabling rectification~\cite{gasse2013}. Eq.~\eqref{quadra} nevertheless suggests that better squeezing can be achieved by using a genuine non-linear system. 

\section{Asymmetric quantum dot}
 We consider a quantum dot for the conductor embedded in the cavity. The situation where the dot is symmetrically coupled to the two leads, discussed in the Supplementary Note 3, is not optimal for squeezing. We thus focus on the asymmetric case where the upper lead is more weakly coupled to the quantum dot than the lower lead. In this case, the voltage drop from the central strip to the ground mainly takes place at the upper dot-lead tunnel contact. The first order cavity-conductor coupling is then of the form $i \lambda \hat{I}_U (\hat{a}^\dagger - \hat{a})$ where $\hat{I}_U$ denotes the electrical current of electrons incoming from the upper lead. The coupling to the lower lead current is neglected.

\changes{In practice, for quantum dot geometries, it may be important to also take into account the coupling of electronic transport to phonons. Following Ref.~\cite{gullans2015}, it would lead to tunneling processes involving the excitation of phonon-photon pairs, degrading the quality of squeezing. Such study is nevertheless beyond the scope of this work and we neglect electron-phonon coupling in what follows.}

The analysis developed above for the tunnel junction can be essentially carried over to the quantum dot, with $\hat{I}_U$ replacing $\hat{I}$ in the Heisenberg-Langevin Eq.~\eqref{langevin}. The noise properties of the quantum dot are derived using scattering theory as discussed in appendix~\ref{appen-landauer}. We retrieve noise factors similar to Eq.~\eqref{noise-terms},
\begin{equation}\label{noise-terms2}
2 S_n (\omega) = \sum_{n'} \left[ c_{n'} c_{n'+n}^* \bar{S}_+ (\omega + e V/\hbar + 2 n' \omega_0) 
 +  c_{n'}^* c_{n'-n} \bar{S}_- (\omega - e V/\hbar - 2 n' \omega_0) \right] e^{-i n \varphi},
\end{equation}
here involving two different  equilibrium noise terms
\begin{equation}\label{noiseterm}
\bar{S}_{\pm} (\omega) = \frac{2 e^2 \Gamma_U \Gamma}{\pi \hbar^2} \int d \varepsilon \frac{f(\varepsilon-\hbar \omega) \left[1-f(\varepsilon) \right]}{(\varepsilon \mp \varepsilon_d)^2 + (\hbar \Gamma/2)^2},
\end{equation}
with the Fermi function $f(\varepsilon) = (1+e^{\varepsilon/k_B T})^{-1}$.
The broadening $\Gamma$ of the dot single energy level, denoted $\varepsilon_d$, can be decomposed according to its coupling to the upper and lower leads $\Gamma = \Gamma_U + \Gamma_L $ with $\Gamma_U \ll \Gamma_L$. The Lorentzian form in the integrand of Eq.~\eqref{noiseterm} describes the Breit-Wigner resonance for transmitting electrons through the  dot~\cite{Rothstein2009}. $\bar{S}_+ (\omega)$ describes electron-hole excitations with energy $\hbar \omega$, where the electron, with energy $\varepsilon$, tunnels from the lower to the upper lead and has to meet the resonance condition of the dot single-level. $\bar{S}_-$ is the same but with hole tunneling. Proceeding with the calculation of cavity properties based on the Heisenberg-Langevin equation of motion, we retrieve the quadrature variances of Eq.~\eqref{quadra} if we define the even/odd parts as $2 \bar{S}_{e/o}(\omega) = \bar{S}_+ (\omega) \pm \bar{S}_- (-\omega)$. 

We have studied numerically the minimization of the variance $\Delta X_1^2$. Quite generally, squeezing optimization requires zero temperature and the four-wave mixing condition $e V=\hbar \omega_0$~\cite{gasse2013}, as well as a sharp resonance, $\Gamma \ll \omega_0$. In this regime, for $\varepsilon_d >0$, $\bar{S}_-$ becomes negligible and $\bar{S}_+$ in Eq.~\eqref{quadra} is either constant for $n > n_{\rm th} = \varepsilon_d/(2 \hbar \omega_0)-1$, or vanishingly small below this threshold. $\bar{S}_e$ and $\bar{S}_o$ simplify in  Eq.~\eqref{quadra} and the summation involves only values of $n$ above the threshold $n_{\rm th}$.

\begin{figure}
\begin{center}
\includegraphics[width=0.5\textwidth]{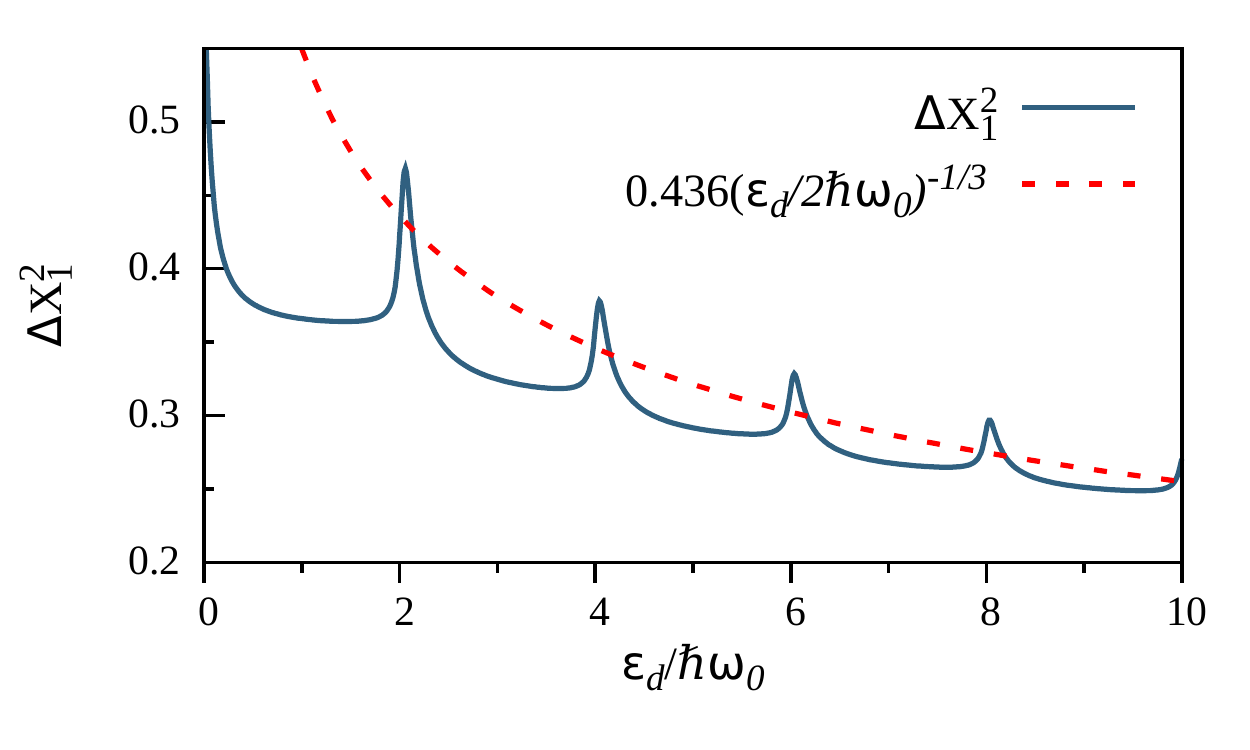}
\caption{Squeezed variance for an asymmetric quantum dot driven by a single-tone excitation $V_{\rm ac}  (t) = V_1 \cos( 2 \omega_0 t + \varphi)$ versus the dot single-level energy, as given by Eq.~\eqref{quadra} and Eq.~\eqref{noiseterm}. $\Delta X_1^2$ is computed at $T=0$, $e V = \hbar \omega_0$ and $\Gamma = 0.1 \omega_0$. For each $\varepsilon_d$, $V_1$ is tuned to minimize  $\Delta X_1^2$. The first local minimum $\Delta X_1^2 = 0.358$ is reached for $\varepsilon_d = 1.54 \, \hbar \omega_0$ and $e V_1/(2 \hbar \omega_0) = 1.16$. Asymptotically at large $\varepsilon_d$, the minima follow the curve $0.436 (\varepsilon_d/2 \hbar \omega_0)^{-1/3}$ (in dotted line) and the optimal ac amplitude follows $e V_1 = \varepsilon_d$. \label{fig-qd}} 
\end{center}
\end{figure}

In the case of a single-frequency ac modulation, Fig.~\ref{fig-qd} shows that the squeezed variance $\Delta X_1^2$ displays a series of local minima, where the values of the  minima decrease with  $\varepsilon_d$. Large single-level energy $\varepsilon_d$ however implies stronger power in the ac excitation signal in order to meet the Breit-Wigner resonance condition. In practice, this requires an ac signal amplitude $V_1$ close to  $\varepsilon_d$ such that electrons can tunnel through  the dot. Perfect squeezing  $\Delta X_1^2=0$ is reached at very large $\varepsilon_d$ but only with a weak power law of coefficient $-1/3$.

Alternatively, a vacuum squeezed state can be reached in the cavity by using a Leviton ac signal with the fundamental frequency $2 \omega_0$. \changes{Leviton pulses were originally~\cite{levitov1996} proposed as voltage excitations designed to transfer a finite number of electrons through a coherent conductor with minimal noise, in analogy with coherent states minimizing quantum-mechanical uncertainty. They consist of sums of Lorentzian pulses with unit flux each. A unit flux represents the attempt to transmit a single electron. Mathematical details are briefly reviewed in appendix~\ref{appen-leviton}. A Leviton can be periodized by having an infinite train of evenly spaced Lorentzian pulses~\cite{Ivanov1997,dubois2013}. Leviton pulses have recently been synthesized and used to perform Hong-Ou-Mandel electronic experiments~\cite{dubois2013b} and electron quantum tomography~\cite{jullien2014}.

The use of a Leviton pulse for squeezing is natural. A Leviton train with periodic short pulses addresses specifically one quadrature (the one to be squeezed), while producing a minimal disturbance (noise) on the quantum conductor.} Taking the limit of a very sharp resonance $\Gamma/\omega_0 \to 0$ \changes{and the bias voltage $e V=\hbar \omega_0$}, one obtains for the squeezed variance
\begin{equation}
\Delta X_1^2 = \frac{\sum_{n=0}^{\infty} ( c_n - c_{n+1} )^2}{\sum_{n=0}^{\infty} ( c_n^2 - c_{n+1}^2 )} = \frac{1-a}{1+a},
\end{equation}
for $0<\varepsilon_d<2 \hbar \omega_0$, and a minimal Heisenberg uncertainty $\Delta X_1 \Delta X_2=1$ \changes{reflecting the minimized perturbation by the Leviton compared to other types of ac excitation. $a=e^{-2 \omega_0 \tau}$ is a parameter related to the width $\tau$ of each Lorentzian pulse.} A Leviton pulse is thus able to produce in optimal conditions an ideal squeezed state with arbitrary compression. 
\begin{figure}
\begin{center}
\includegraphics[width=0.5\textwidth]{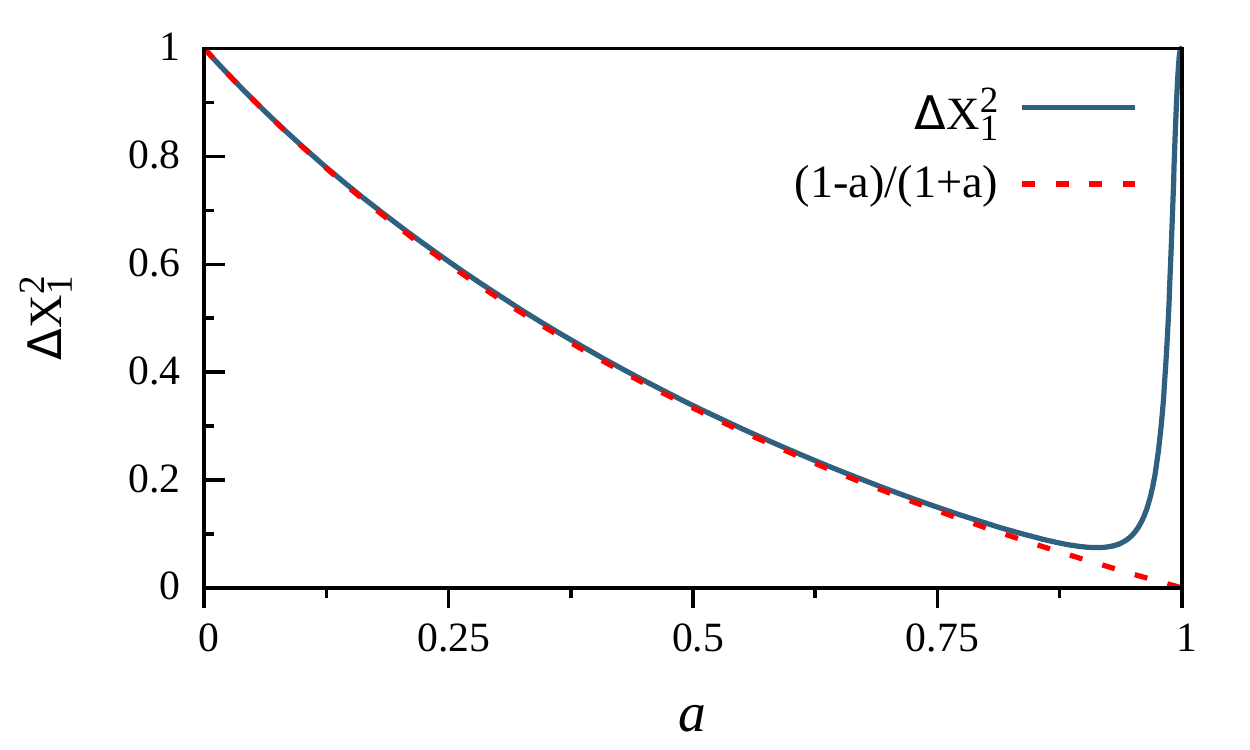}
\caption{Squeezed variance for the asymmetric quantum dot as a function of the Leviton pulse excitation parameter $a$. The continuous line was computed using Eq.~\eqref{quadra} and Eq.~\eqref{noiseterm} at zero temperature, $e V = \hbar \omega_0$,  $\Gamma/\omega_0 = 0.1$, $\varepsilon_d = 6 \, \hbar \omega_0$ and the Leviton $c_n$ coefficients (see appendix~\ref{appen-leviton}). It is compared to the asymptotical expression $(1-a)/(1+a)$ (dotted line) valid for an infinitely sharp resonance $\Gamma \to 0$. Note that the limit $a=1$ corresponds to a dc voltage $e V = - \hbar \omega_0$ producing no squeezing $\Delta X_1^2 =1$, indicating that the limits $\Gamma/\omega_0 \to 0$ and $a\to 1$ do not commute. \label{fig-leviton}} 
\end{center}
\end{figure}

As shown in Fig.~\ref{fig-leviton}, squeezing saturates at finite damping $\Gamma$. For a reasonable damping rate $\Gamma/\omega_0 = 0.1$, we find that the squeezed variance can still be reduced down to $\Delta X_1^2 = 0.075$ for $a=0.91$. A Leviton excitation with parameter $a$ close to $1$ exhibits \changes{sharp voltage pulses of width $\tau \simeq (1-a)/(2 \omega_0)$.} Similarly to the tunnel junction, it corresponds to concentrate short-time pulses of high voltage when the amplitude of the squeezed quadrature is maximum while the other quadrature vanishes. 

\section{Cavity Readout}
 Cavity squeezing could be evidenced by an in-situ measurement using a  qubit and its anisotropic radiative decay~\cite{murch2013}. It is also possible to demonstrate squeezing in the cavity by measuring the output field. So far, we have neglected in our discussion the coupling of the cavity to the external electromagnetic modes, see Fig.~\ref{fig1}. They introduce an additional damping rate $\kappa_0$. As detailed in the Supplementary Note 2, this external damping adds two terms in the Heisenberg-Langevin Eq.~\eqref{langevin}, which now reads
\begin{equation}\label{langevin-mod}
\dot{\hat{a}} + i \omega_0 \hat{a} + \frac{ \kappa+\kappa_0}{2} \hat{a} = - \sqrt{\kappa_0} \, \hat{b}_{\rm in} (t) + \lambda  \hat{I} (t).
\end{equation}
$\hat{b}_{\rm in} (t)$ is the input field, it describes the quantum state of incident photons on the cavity. In the absence of an input drive, it corresponds to vacuum noise  with $\langle \hat{b}_{\rm in} (t) \hat{b}_{\rm in}^\dagger (t') \rangle = \delta (t-t')$ at zero temperature. The output field, describing photons escaping the cavity, is given by $\hat{b}_{\rm out} =  \hat{b}_{\rm in} + \sqrt{\kappa_0} \, \hat{a}$ and obtained by solving Eq.~\eqref{langevin-mod}. 

Squeezing properties of the cavity are revealed by homodyne detection of the output field~\cite{yurke1988,castellanos2008}, mixing it with a local oscillator with the cavity frequency $\omega_0$ and phase $\theta$. The power spectrum $S_{ D} (\omega)$~\cite{clerk-schoelkopf-rmp2010} of the homodyne detector signal $\hat{I}_D (t) \propto \hat{b}_{\rm out}  (t)e^{i (\omega_0 t+\theta)} + h.c.$ exhibits a deep at zero frequency and $\theta = \pi/2 + \varphi/2$, where the squeezing effect is most visible, with
\begin{equation}
\frac{S_{ D} (\omega=0)}{S_{ D}^0} = 1 +  \frac{4 \kappa_0}{\kappa} \left( \Delta X_1^2 -1 \right) < 1,
\end{equation}
assuming weak external damping $\kappa_0 \ll \kappa$. $S_{D}^0$ denotes the vacuum level, measured in the absence of ac excitation. A value smaller than one in this equation indicates a squeezed output field. Interestingly, output squeezing is weak in this limit due to pollution by the input vacuum noise. As discussed in appendix~\ref{damping}, the output field gets more squeezed as $\kappa_0$ increases, the best squeezing being obtained for equal external and electronic dampings. This is however at the price of weaker squeezing in the cavity. Perfect squeezing could even be reached in the output field, limiting in this case the cavity state to half-squeezing.



\section{Conclusion and outlook}
 We studied the squeezing generated in a resonant cavity by coupling it to a mesoscopic conductor under parametric excitation. We showed that the quality of squeezing can be improved by enhancing nonlinearities in the conductor and by concentrating the voltage excitation pulses at instants where the squeezed quadrature amplitude reaches its maximum. In optimal conditions, perfect squeezing can even be achieved. We remark that nonlinearities could also be enhanced in a tunnel junction by increasing the coupling to the cavity and the associated dynamical Coulomb blockade~\cite{Parlavecchio2015}. More generally, our results can be easily extended to other quantum conductors for which the photo-assisted noise spectra are known or can be computed. They also suggest the possibility to engineer squeezed light for quantum information using electronic quantum conductors.

{\it Acknowledgments.} We thank C. Altimiras, B. Huard, P. Joyez, F. Mallet, F. Portier, P. Simon and M. Trif for fruitful discussions. U.C.M. acknowledges the support from CNPq-Brazil (Project No. 229659/2013-6). \\

\appendix

\section{Leviton excitation}\label{appen-leviton}

A Leviton is a pulse shape designed to minimize shot noise in driving electric current~\cite{levitov1996}. \changes{In its periodized form, the time-dependent voltage is given by a sum of quantized Lorentzian pulses
\begin{equation}\label{pulse-shape}
V_{\rm ac} (t) = -2 \hbar \omega_0 + \frac{h}{e} \sum_{l \in \mathbb{Z}} \frac{1}{ \pi} \frac{\tau}{\tau^2+(t- l \pi/\omega_0)^2} ,
\end{equation}
of common width $\tau$. The ac phase is obtained by time-integration, $\phi_{\rm ac} (t) =(e/\hbar) \int^t d t' V_{\rm ac} (t')$. It} is conveniently written using the cyclic variable $z=e^{2 i \omega_0 t}$, namely after a few algebraic manipulation~\cite{Ivanov1997}
\begin{equation}\label{leviton}
e^{i \phi_{\rm ac}} = \frac{1}{z} \frac{z-a}{1- a z}
\end{equation}
\changes{where $0<a=e^{-2 \omega_0 \tau}<1$ is related to the pulses width. The conventional Leviton pulse, shaped to minimize zero-frequency shot noise, has the dc level $e V= 2 \hbar \omega_0$, cancelling the first term in Eq.~\eqref{pulse-shape}. In the main text, the dc bias chosen to optimize squeezing is $e V= \hbar \omega_0$. It can be understood by noting that the goal here is to minimize finite frequency noise correlators at the frequency  $\omega_0$.} Expanding Eq.~\eqref{leviton} in powers of $z$, we obtain $c_{-1} = -a$, $c_{n\ge0} = a^n (1-a^2)$: there is a strong imbalance between absorption and emission of photons. $a=1$ is no longer an ac modulation but corresponds to a shift of the dc voltage  by $-2 (\hbar/e) \omega_0$.

\section{Keldysh formulation and Heisenberg-Langevin equation}\label{keldysh}

The out-of-equilibrium physics of our system is conveniently described within the Keldysh path-integral formalism~\cite{torre2013}, enabling a systematic integration of the electronic degrees of freedom and yielding an effective action for the photons~\cite{clerk2004}. The action obtained, Eqs.~\eqref{seff} and~\eqref{sa}, can be shown to be equivalent to the Heisenberg-Langevin equation of motion Eq.~\eqref{langevin} used in the main text, in the limit of small $\kappa/\omega_0 \ll 1$. For a direct derivation of the Heisenberg-Langevin equation in the spirit of input-output theory, see Supplementary Note 2.

The partition function (hereafter $\hbar=1$) 
\begin{equation}
{\cal Z} = \int {\cal D}[a,a^*] e^{i S_{\rm cav}} \int {\cal D}[c,c^*] e^{i (S_{\rm e} + S_{\rm ep})}
\end{equation}
involves an integration over complex-valued fields, $a$,$a^*$ for photons and $c$,$c^*$ for electrons. $S_{\rm e}$ is the action for the isolated conductor, $S_{\rm cav}$ the photonic part corresponding to the Hamiltonian $H_{\rm cav} = \hbar \omega_0 \hat{a}^\dagger \hat{a}$ and the electron-photon coupling is to first order
\begin{equation}
S_{\rm ep} = -\frac{i \lambda}{\sqrt{2}} \int_{-\infty}^{+\infty} d t  \sum_{\eta \pm 1} \eta \, I_{\eta} (t) (a^*_{\eta} (t) - a_{\eta} (t)) 
\end{equation}
where $\eta$ denotes the Keldysh time branch. $I$ is the quantum conductor current written with complex-valued fields.

The electronic part can be rigorously integrated using the cumulant expansion $\langle e^{i S_{\rm ep}} \rangle_e = e^{i  \langle S_{\rm ep} \rangle_e -(1/2) \langle \delta S_{\rm ep}^2 \rangle_e +\ldots}$, where we use the notations $\delta S_{\rm ep} = S_{\rm ep} - \langle S_{\rm ep} \rangle$ and  $\int {\cal D}[c,c^*] e^{i S_{\rm e}} A =  \langle A \rangle_e$. We assume $\langle S_{\rm ep} \rangle_e =0$, a finite $\langle S_{\rm ep} \rangle_e$ can be absorbed by a small shift $\propto g$ in the cavity fields. 
To summarize, we square $S_{\rm ep}$, take its quantum average restricted to electronic degrees of freedom, and thus obtain a self-energy kernel for the photons involving current noise correlators. For clarity, we switch to classical/quantum variables, $a_{cl/q} = (a_+ \pm a_-)/\sqrt{2}$ and write the action in frequency space in order to take advantage of the current noise correlators given in Eq.~\eqref{noise}. Summing $S_{\rm cav}$ and $(i/2) \langle S_{\rm ep}^2 \rangle_e$, we find the effective action
\begin{equation}\label{seff}
S_{\rm eff} = \int_\omega \begin{pmatrix} a^*_{cl} & a_q^*  \end{pmatrix}
\begin{pmatrix} 0 & G_A^{-1} (\omega) \\ G_R^{-1} (\omega) & - \Sigma_K  \end{pmatrix} \begin{pmatrix} a_{cl} \\ a_q 
\end{pmatrix} + S_a,
\end{equation}
where we used the notation $\int_\omega = \int_{-\infty}^{+\infty} \frac{d \omega}{2 \pi}$, and $a_{cl} (t) = \int_\omega a_{cl} (\omega) e^{-i \omega t}$. The retarded photon Green's function $G_R^{-1} (\omega) = \omega -\omega_0 + i \kappa/2$  has a pole shifted by half the damping rate $\kappa$ (see main text). The quantum-quantum self-energy part is $\Sigma_K = -i \lambda^2 [ S_0(\omega_0)+S_0(-\omega_0)]$. The effective action $S_{\rm eff}$ includes also an anomalous term, responsible for state squeezing,
\begin{equation}\label{sa}
S_a = - i \lambda^2 \int_\omega  \left( a^*_q(\omega) a^*_q(2 \omega_0 - \omega) S_1 (\omega_0) + {\rm c.c.} \right).
\end{equation}
Note that the real part of the self-energy induces \changes{in general} a cavity pull which has been  absorbed into a redefinition of $\omega_0$. \changes{Computing this frequency shift consistently requires the second order term in the expansion of $H_T$ in powers of  $g/\omega_0$, namely 
\begin{equation}
\delta S_{\rm ep}^{(2)} = -\frac{e^2 \lambda^2}{2} \int_{-\infty}^{+\infty} d t  \sum_{\eta \pm 1} \eta \, V_{T,\eta} (t)  \left[ a^*_{\eta} (t) - a_{\eta} (t) \right]^2,
\end{equation}
where $V_T = {\cal T} +  {\cal T}^\dagger$. The cavity frequency shift then vanishes for a tunnel junction.}

The derivation of Eqs.~\eqref{seff} and~\eqref{sa} relies on the rotating-wave approximation, valid for $\kappa \ll \omega_0$, where the fields $a_{cl/q} (\omega), a_{cl/q}^* (\omega)$ take significant values only for $\omega \simeq \omega_0$. \changes{In principle, the anomalous part of the action  $S_a$  also contains terms with $a_{cl} (\omega) a_q(2 \omega_0 - \omega)$, corresponding to the effect of non-stationary noise terms on the damping. Those terms are found to be proportional to $S_1 (\omega) - S_1(2\omega_0 -\omega)$ and thus vanish for  $\omega \simeq \omega_0$, with the small parameter $\kappa/\omega_0$.

Finally, we discuss the connexion between the quadratic action of Eq.~\eqref{seff} and the Heisenberg-Langevin equation~\eqref{langevin}. Quite generally, it is known that current fluctuations in a tunnel junction, or a quantum conductor, are not gaussian. Nevertheless, computing the non-gaussian current contributions to the statistics of photons, one finds that they are small compared to the dominant Wick-like contractions among the current operators. This is true in the limit of weak damping, $\kappa \ll \omega_0$, where cavity correlation functions only involve current operators $\hat{I}$ at frequencies $\pm \omega_0$. For example, the fourth-order cavity field correlator gives, for $|t_i| \ll 1/\omega_0$, 
\begin{equation}
\langle \hat{a}^{(\dagger)}_{t_1}  \hat{a}^{(\dagger)}_{t_2}  \hat{a}^{(\dagger)}_{t_3}  \hat{a}^{(\dagger)}_{t_4} \rangle = \left( \sum_{P \in \mathbb{S}_4}  \langle \hat{a}^{(\dagger)}_{t_{P(1)}}  \hat{a}^{(\dagger)}_{t_{P(2)}} \rangle  \langle \hat{a}^{(\dagger)}_{t_{P(3)}}  \hat{a}^{(\dagger)}_{t_{P(4)}} \rangle \right) 
\left( 1 + {\cal O} \left( \frac{\kappa}{{\rm max} ( |e V/\hbar - \omega_0|,e V_{\rm ac}/\hbar )} \right) \right),
\end{equation}}
\changes{implying photon gaussian statistics except for the specific case of $e V$ close to $\hbar \omega_0$ with no ac excitation. A related discussion can be found in Refs.~\cite{beenakker2001,beenakker2004}. The reason is that an electron-hole excitation with energy $\hbar \omega_0$ created by a current operator $\hat{I}_{\omega_0}$ must be destroyed by another single current operator, the phase-space for alternative processes - where electron and hole are annihilated by two distinct current operators - being negligible for weak damping $\kappa$. This argument pertains to higher-order correlation functions such that, for the purpose of photon statistics, \changes{it is legitimate to keep only the gaussian part of electronic current fluctuations. The resulting cavity field statistics are obviously gaussian.}

The above cumulant expansion can be rigorously stopped after the second order and the  gaussian action in Eq.~\eqref{seff} becomes exact as long as $\kappa$ is negligible with respect to $\omega_0$. Computing second-order cavity correlations functions, with different ordering of $\hat{a}$ and $\hat{a}^\dagger$, we find coinciding results for the gaussian action and the Heisenberg-Langevin evolution. This completes the proof of the equivalence of the two formulations. This comparison differs from the standard derivation of a classical Langevin equation using the Keldysh action~\cite{kamenev2011}, in which case information about operator ordering is lost.
}


\section{Optimized squeezing for a tunnel junction}\label{appen-minimum}

In this section, we set  $\hbar=1$ for simplicity.
We focus on the zero-temperature case, relevant to maximize the cavity state compression. In this case, 
\begin{equation}
\bar{S}_o \left[e V + (2 n+1) \omega_0 \right] = \frac{1}{R_T} \Big[ e V  + (2 n+1) \omega_0 \Big]
\end{equation}
 and the denominator in Eq.~\eqref{quadra} of the main text simplifies to
\begin{equation}
\sum_{n \in \mathbb{Z}} \left( | c_n |^2 - |c_{n+1} |^2 \right)  \Big( e V  + (2 n+1) \omega_0 \Big) =2 \omega_0,
\end{equation}
regardless of $V$ and the $c_n$ coefficients. At the optimal dc voltage $e V = \hbar \omega_0$, the squeezed variance takes the simple form
\begin{equation}\label{var}
\Delta X_1^2 = \sum_{n \in \mathbb{Z}} | c_n \pm c_{n-1} |^2 |n|,
\end{equation}
which we still need to minimize with respect to the distribution of Fourier coefficients $c_n$,
\begin{equation}\label{fourier}
c_n = \frac{\omega_0}{\pi} \int_0^{\pi/\omega_0} d t \, e^{i  \phi_{\rm ac} (t)} e^{-2 i n \omega_0 t}.
\end{equation}
We now prove that the pulse shape of Eq.~\eqref{pulse} in the main text, corresponding to the piecewise linear phase $\phi_{\rm ac,opt}  (t) = \pi/2-\omega_0 t$, extremizes the variance $\Delta X_1^2$. We first differentiate Eq.~\eqref{fourier} to obtain
\begin{equation}
\frac{\partial c_n}{\partial \phi_{\rm ac} (t)} = \frac{i \omega_0}{\pi} e^{i \phi_{\rm ac} (t)} e^{-2 i n \omega_0 t},
\end{equation}
which gives $-(\omega_0/\pi) e^{-i (2n+1) \omega_0 t}$ when evaluated at $\phi_{\rm ac,opt}  (t)$. Using this result, we can proceed with the derivative of $\Delta X_1^2$ with respect to an arbitrary form of $\phi_{\rm ac} (t)$
\begin{equation}
\frac{\partial \Delta X_1^2}{\partial \phi_{\rm ac} (t)} = 2 \sum_n |n| {\rm Re} \left[ (c_n^* - c_{n-1}^*) (\partial_{\phi_{\rm ac}} c_n - \partial_{\phi_{\rm ac}} c_{n-1}) \right].
\end{equation}
We evaluate this derivative with $\phi_{\rm ac,opt}  (t)$ and its coefficients
\begin{equation}\label{coeff}
c_n = \frac{1}{\pi} \frac{1}{n+1/2},
\end{equation}
and obtain
\begin{equation}
 \frac{\partial \Delta X_1^2}{\partial \phi_{\rm ac} (t)} = \frac{2 \omega_0}{\pi} {\rm Re} \Bigg[  \sum_{n \in \mathbb{Z}} \frac{|n|}{(n+1/2) (n-1/2)}  
  \left( e^{-i (2 n+1)\omega_0 t} - e^{-i (2 n-1)\omega_0 t} \right) \Bigg] = 0,
\end{equation} 
which completes the proof. 

Inserting the coefficients Eq.~\eqref{coeff} into the quadrature variance Eq.~\eqref{var}, we find
\begin{equation}
\Delta X_1^2 = \sum_{n \in \mathbb{Z}} \frac{|n|}{\pi^2 (n+1/2)^2 (n-1/2)^2} = \frac{4}{\pi^2}.
\end{equation}
We also checked numerically that $\phi_{\rm ac,opt}  (t)$ reaches the global minimum of $\Delta X_1^2$.

\section{Landauer-B\"uttiker calculation of the noise}\label{appen-landauer}

The noise properties of the quantum dot are derived using the scattering, or Landauer-B\"uttiker, formalism~\cite{pedersen1998}. The current operator is expanded over the basis of one-particle scattering states originating from both leads. The general case is reviewed in the Supplementary Note 4, we focus here on the asymmetric case where the probability for single-electron transmission at resonance $4 \Gamma_U/\Gamma \ll 1$ is small, and expressions simplify. Omitting spin, the current operator has the form
\begin{equation}\label{currentasym}
\begin{split}
\hat{I}_U (t) = \frac{i e \sqrt{\Gamma_U \Gamma}}{h} 
\int \int d \varepsilon_1 d \varepsilon_2  \, \hat{c}_U^\dagger (\varepsilon_1) g (\varepsilon_2) \hat{c}_L (\varepsilon_2) \\[1mm] \times e^{i [(\varepsilon_1 - \varepsilon_2 +V)t /\hbar+\phi_{\rm ac} (t)]} 
 + h.c.
\end{split}
\end{equation}
with the Breit-Wigner resonant function $g(\varepsilon) = (\varepsilon - \varepsilon_d + i \hbar \Gamma/2)^{-1}$. The operator  $\hat{c}_{U/L} (\varepsilon)$ annihilates an electron in a scattering state of energy $\varepsilon$ incoming from the upper/lower lead. The normalisation is fixed by the average 
\begin{equation}
\langle \hat{c}^\dagger_\alpha (\varepsilon) \hat{c}_{\alpha'} (\varepsilon') \rangle = \delta_{\alpha,\alpha'} f(\varepsilon) \delta ( \varepsilon - \varepsilon')
\end{equation}
with the Fermi function $f(\varepsilon) = (1+e^{\varepsilon/k_B T})^{-1}$. Due to the small capacitance at the upper dot-lead contact, the dc and ac bias voltages are applied essentially across this tunnel contact, the voltage potentials on both the quantum dot and the lower lead are fixed to the ground. Apart from the Breit-Wigner function, the rest of the calculation is similar to the case of a tunnel junction. The two-current correlators have the form of Eq.~\eqref{noise} and Eq.~\eqref{noise-terms}, where the equilibrium noise terms are given Eq.~\eqref{noiseterm} in the main text.

\section{Finite external damping}\label{damping}

We briefly discuss the case of a bare cavity damping $\kappa_0$ comparable to the electronic damping $\kappa$, but still much smaller than the resonator frequency $\omega_0$. The complete Heisenberg-Langevin equation~\eqref{langevin-mod} is solved by considering both the input field and electronic current fluctuations. One obtains for the two cavity field quadratures
\begin{equation}
\Delta X_{1/2}^2 (\kappa_0) = 1 + \frac{\kappa}{\kappa + \kappa_0} \left( \Delta X_{1/2}^2 (0) -1 \right),
\end{equation}
where 
\begin{equation}
\Delta X_{1/2}^2 (0) = \frac{S_0 (\omega_0) + S_0 (-\omega_0) \mp 2 {\rm Re}  \left[ S_1 (\omega_0)\right]}{S_0 (\omega_0) - S_0 (-\omega_0)}
\end{equation}
denote their variances in the absence of intrinsic damping $\kappa_0$, also given by Eq.~\eqref{quadra} and discussed in length in the main text. Additionally, one finds for the output field squeezing, characterized by the power spectrum,
\begin{equation}
\frac{S_{ D} (\omega=0)}{S_{ D}^0}  = 1 +  \frac{4 \kappa_0 \kappa}{(\kappa + \kappa_0)^2} \left( \Delta X_1^2 (0) -1 \right).
\end{equation}
Whereas a vanishing $\Delta X_1^2 (0)$ clearly optimizes squeezing in both the cavity and output fields, there is no such choice for $\kappa_0$. Increasing $\kappa_0$ from zero improves squeezing in the output field but degrades cavity squeezing. Perfect squeezing in the output field is reached for $\kappa = \kappa_0$, with vanishing $\Delta X_1^2 (0)$, in which case the cavity field is only half-squeezed.


\clearpage

\section*{Cavity squeezing by a quantum conductor (Supplementary material)}

\renewcommand{\thesection}{S-\Roman{section}}
\renewcommand{\theequation}{S-\arabic{equation}}
\renewcommand{\thefigure}{S-\arabic{figure}}
\renewcommand{\appendixname}{}

\setcounter{equation}{0}
\setcounter{section}{0}

Cited equations not preceded by $S-$ refer to the main text.
Unless indicated otherwise, we set $\hbar=1$ throughout the Supplementary material.

\section{Current-current correlation function}

In this section, we derive the current-current correlation function of an ac driven tunnel junction in the absence of cavity under the general form of Eq.~\eqref{noise} in the main text. We use the definition
\begin{equation} \label{cur-corr}
\langle \hat{I}(\omega_{1})  \hat{I}(\omega_{2})  \rangle = \int_{-\infty}^{\infty} dt_{1}dt_{2} \langle \hat{I}(t_{1})  \hat{I}(t_{2})  \rangle e^{i\omega_{1} t_{1}} e^{i\omega_{2} t_{2}}.
\end{equation}
 The current operator through the junction is given by
\begin{equation} \label{current}
\hat{I}(t) = i \sum_{\kq} \left( \gamma_{\kq} e^{i [V t + \phi_{\rm ac}(t)]}\hat{c}_{\kl}^{\da}(t)\hat{c}_{\kr}(t) - \rm{h.c.} \right),
\end{equation}
corresponding to the operator 
\begin{equation}
{\cal T}^\dagger = \sum_{\kq}\hat{c}_{\kl}^{\da} \hat{c}_{\kr}
\end{equation} 
in Eq.~\eqref{eq-tunn} of the main text. Here $\hat{c}_{\mathbf{k}\alpha}^{\da}$ ($\hat{c}_{\mathbf{k}\alpha}$) is the electron creation (annihilation) operator in lead $\alpha=\{\rm{U,L}\}$, where $U$ and $L$ stand for upper and lower leads in a top-down geometry. In the absence of interaction and cavity photons, the Heisenberg time evolution of fermion operators is simply $\hat{c}_{\kr}(t)=\hat{c}_{\kr}e^{-i \varepsilon_{\mathbf{q}}t}$. Using  the expansion in Fourier coefficients
\begin{equation}
e^{i \phi_{\rm ac} (t)} = \sum_{n \in \mathbb{Z}} c_n e^{2 i n  \omega_0 t} e^{i n \varphi},
\end{equation}
the two-time correlation function takes the form
\begin{align}  \label{cur-corr-time}
& \langle \hat{I}(t_{1})   \hat{I}(t_{2})\rangle  = \sum_{\substack{\kq \\  n_{1},n_{2} }}|\gamma_{\kq}|^2 \bigg( c_{n_{1}}c_{n_{2}}^{*}  f(\varepsilon_{\mathbf{k}})\bar{f}(\varepsilon_{\mathbf{q}})  \\
& \times  e^{i(\omega_{\kq}+eV)(t_{1}-t_{2})}e^{2i\omega_{0}(n_{1}t_{1}-n_{2}t_{2})} + \rm{c.c.} \bigg),  \nonumber
\end{align}
where we incorporated the phase terms in the Fourier coefficients $c_n e^{i n \varphi} \to c_n$ and defined $\omega_{\kq}=\varepsilon_{\mathbf{k}}-\varepsilon_{\mathbf{q}}$.  $f(\varepsilon) = (1+e^{\beta \varepsilon})^{-1}$ and $\bar{f} (\varepsilon)=1-f(\varepsilon)$ are the electron and hole Fermi-Dirac distribution functions.
 Substituting Eq.~\eqref{cur-corr-time} in Eq.~\eqref{cur-corr} and integrating over $t_{1}$ and $t_{2}$, we obtain  
\begin{align}  \label{cur-corr-1}
\langle & \hat{I}(\omega_{1})  \hat{I}\omega_{2}) \rangle  = \frac{2\pi}{R_{\rm T}} \sum_{n_{1},n_{2}} \int d\varepsilon_k d\varepsilon_q \Big[ c_{n_{1}}c_{n_{2}}^{*} f(\varepsilon_k)\bar{f}(\varepsilon_q) \nonumber   \\
& \times \delta(\omega_{kq}+ \omega_{1}+  eV + 2n_{1}\omega_{0} )\delta(\omega_{1}+\omega_{2} -2(n_{2}-n_{1})\omega_{0})  \nonumber \\
& + c_{n_{1}}^{*}c_{n_{2}} f(\varepsilon_k)\bar{f}(\varepsilon_q) \delta(\omega_{kq}+ \omega_{1}- eV - 2n_{1}\omega_{0})  \nonumber \\
& \times \delta(\omega_{1}+\omega_{2} -2(n_{1}-n_{2})\omega_{0})  \Big].
\end{align}
In deriving this expression, we  used the standard assumptions of weak momentum dependence of $\gamma_{\kq}$ and of the density of state in the leads, taken to be constant and denoted $\nu_0$. We introduced the tunnel resistance $R_{T}=1/(2\pi |\gamma|^2 \nu_{0}^{2})$. Before pursuing our derivation of current-current correlation function, we define the unsymmetrized equilibrium Johnson-Nyquist noise for a tunnel junction, in absence of both dc and ac bias, as
\begin{equation}
\bar{S}(\omega) = \int_{-\infty}^{\infty} \langle I(t)I(0)\rangle e^{i\omega t}. 
\end{equation}
Using the expression Eq.~\eqref{current} of the current operator, we readily find
\begin{align} \label{john-nyq-noise}
\bar{S}(\omega)&=\frac{2}{ R_{\rm T}} \int d\varepsilon_k d\varepsilon_q f(\varepsilon_k)\bar{f}(\varepsilon_q)\delta(\omega_{\kq}+\omega) \nonumber \\
&=\frac{1}{R_{\rm T}}\frac{2\omega}{(1-e^{-\omega/k_{\rm B}T})}.
\end{align}
Using the above definition, we are able to rewrite the current-current correlator [Eq. \eqref{cur-corr-1}] as
\begin{align}  \label{cur-corr-2}
\langle & \hat{I}(\omega_{1})  \hat{I}\omega_{2}) \rangle = \sum_{n}S_{n}(\omega_{1})2\pi\delta(\omega_{1}+\omega_{2} -2n\omega_{0}) ,
\end{align}
where the different noise terms are defined as
\begin{align}  \label{noise-1}
S_{n}(\omega_{1}) &= \frac{1}{2}\sum_{n_{1}} \Big[c_{n^{\prime}}c_{n^{\prime}+n}^{*} \bar{S}(\omega_{1}+  eV + 2n^{\prime}\omega_{0}) \nonumber \\
& + c_{n^{\prime}}^{*}c_{n^{\prime}-n}\bar{S}(\omega_{1}-eV - 2n^{\prime}\omega_{0}) \Big].
\end{align}
We thus obtain Eqs.~\eqref{noise} and \eqref{noise-terms} of the main text once the phase terms $c_n \to c_n e^{i n \varphi} $ have been reinstated.

\section{Heisenberg-Langevin equation}

We derive in this section the Heisenberg-Langevin equation, Eq.~\eqref{langevin} in the main text, following the steps of standard input-output theory~\cite{clerk-schoelkopf-rmp2010}. A similar analysis can be found in Ref.~\cite{dmytruk2015}. The cavity is coupled to a quantum conductor and also to external electromagnetic modes. Both play the role of an environment for the cavity, absorbing and emitting photons via quantum vacuum noise terms. The Hamiltonian of the complete system is 
\begin{align} \label{Ham-tot}
H &=  \omega_{0} \hat{a}^{\da}\hat{a} + \sum_{\ki} \varepsilon_{\ki}\hat{c}_{\ki}^{\da}\hat{c}_{\ki} + i\lambda (\hat{a}^{\da}-\hat{a})\hat{I} \nonumber \\
&+ \sum_{n}\Omega_{n}\hat{b}_{n}^{\da}\hat{b}_{n} -i\sum_{n}g_{n}(\hat{a}^{\da}\hat{b}_{n}-\hat{a}\hat{b}_{n}^{\da}).
\end{align}
The first term describes the isolated cavity, the second one the scattering states $\hat{c}_{\ki}$~\cite{blanter2000}  incoming from both leads $\alpha=U/L$, and which includes scattering by the quantum conductor. The third term is the coupling between the cavity single-mode and the electric current through the conductor. For a tunnel junction, the scattering states  $\hat{c}_{\ki}$ coincide with lead electrons to lowest order in the tunneling. The current operator $\hat{I}$ in the Hamiltonian is then
\begin{equation}
\hat{I} = i \sum_{\kq} \left( \gamma_{\kq} e^{i [V t + \phi_{\rm ac}(t)]}\hat{c}_{\kl}^{\da} \hat{c}_{\kr} - \rm{h.c.} \right).
\end{equation}
It is in the Schr\"odinger picture where the only time dependence is through the voltage.  In the rest of this section, we will focus our analysis on the tunnel junction for simplicity.

The last two terms in Eq.~\eqref{Ham-tot} stand for the external photonic modes whose coupling to the cavity is assumed to be weak. We now switch to the Heisenberg picture where $\hat{a} (t) = e^{i H t} \hat{a} e^{-i H t}$ and compute the equation of motion (EOM) for the cavity field from $\dot{\hat{a}}(t) = i [H,\hat{a}] (t)$. The external modes can be treated perturbatively, as detailed in Ref.~\cite{clerk-schoelkopf-rmp2010}, reducing the EOM to
\begin{equation} \label{eom-cav}
\dot{\hat{a}}(t) = -i\omega_{0}\hat{a}(t)- \frac{\kappa_{0}}{2} \hat{a}(t)-\sqrt{\kappa_{0}} \, \hat{b}_{\rm in}(t) + \lambda \hat{I}_{\rm H}(t),
\end{equation}
where $\kappa_0$ denotes the cavity damping rate due to these modes. $\hat{b}_{\rm in}(t)$ is an input field which feeds the cavity with vacuum noise at zero temperature.  The cavity field is also coupled to the output field (from which cavity properties can be extracted) through the boundary condition  
\begin{equation}  \label{out-cav}
\hat{b}_{\text{out}} (t) = \hat{b}_{\rm in} (t)+\sqrt{\kappa_{0}} \hat{a}(t).
\end{equation}
$\hat{I}_{\rm H}(t)$ in Eq.~\eqref{eom-cav} denotes the current in the Heisenberg picture. It is evolved with the complete Hamiltonian $H$ including, via coupling to the cavity, the cavity single-mode and its electromagnetic environment. We will write the time evolution of $\hat{I}_{\rm H}(t)$ perturbatively in the cavity-conductor coupling $\lambda$ and insert the result into the cavity EOM.

Let us start by introducing the Heisenberg-picture operators
\begin{equation} \label{curr-op}
\hat{f}_{\kq}^{\rm H}(t) = \hat{c}_{\kr}^{\da}(t)\hat{c}_{\kl}(t)
\end{equation}
which appear in the current $\hat{I}_{\rm H}(t)$. Taking the derivative with respect to time, one obtains 
\begin{align} \label{curr-op-eom}
\dot{\hat{f}}_{\kq}^{\rm H}(t) &=-i(\varepsilon_{\kl}-\varepsilon_{\kr})\hat{f}_{\kq}^{\rm H}(t) - \lambda[\hat{V}_{\rm int}(t),\hat{f}_{\kq}^{\rm H}(t)]
\end{align}
in which $\hat{V}_{\rm int}(t) = [\hat{a}^{\da}(t)-\hat{a}(t)]\hat{I}_{\rm H}(t)$. Assuming a weak cavity-conductor coupling, we solve Eq.~\eqref{curr-op-eom} perturbatively. The lowest order in $\lambda$ is readily integrated to give
\begin{equation} \label{curr-op-zero}
\hat{f}_{\kq}^{\rm H}(t)= \hat{f}_{\kq}e^{-i\omega_{\kq}t},
\end{equation}
where we define $\omega_{\kq} = \varepsilon_{\kl}- \varepsilon_{\kr}$. At this order, the time evolution does not include the cavity mode and the current operator is given by Eq.~\eqref{current} and is denoted $\hat{I}(t)$. Replacing $\hat{f}_{\kq}^{\rm H}(t)$ in the last term of Eq.~\eqref{curr-op-eom} by the zeroth order solution Eq.~\eqref{curr-op-zero}, and the current $\hat{I}_{\rm H}(t)$ by $\hat{I} (t)$ in $\hat{V}_{\rm int}(t)$, we obtain the first order correction
\begin{align} \label{curr-op-first}
\dot{\hat{f}}_{\kq}^{\rm H}(t) &= -i\omega_{\kq} \hat{f}_{\kq}^{\rm H} -\lambda[\hat{V}_{\rm int}(t),\hat{f}_{\kq}]e^{-i\omega_{\kq}t}.
\end{align}
This differential equation can be integrated with the result
\begin{align} \label{curr-op-sol}
\hat{f}_{\kq}^{\rm H}(t) &= \hat{f}_{\kq}e^{-i\omega_{\kq}t} -\lambda \int_{-\infty}^{t}d\tau [\hat{V}_{\rm int}(\tau),\hat{f}_{\kq}]e^{-i\omega_{\kq}t} .
\end{align}
We consider the current and obtain
\begin{align} \label{curr-op-final}
\hat{I}_{\rm H}(t) &= i  \sum_{\kq}( \gamma_{\kq} e^{i [V t + \phi_{\rm ac}(t)]} \hat{f}_{\kq}^{\rm H \da}(t)-  \rm{h.c.}) \nonumber \\
&=\hat{I}(t)-\lambda \int_{-\infty}^{t}[a^{\da}(\tau)-a(\tau)][\hat{I}(\tau),\hat{I}(t)] d\tau.
\end{align}
Substituting Eq. \eqref{curr-op-final} in Eq. \eqref{eom-cav} we obtain 
\begin{align} \label{eom-cav-1}
\dot{\hat{a}}(t) &= -i\omega_{0}\hat{a}(t)- \frac{\kappa_{0}}{2} \hat{a}(t) -\sqrt{\kappa_{0}} \, \hat{b}_{\rm in}(t) + \lambda \hat{I}(t) \nonumber \\
&-\lambda^{2} \int_{-\infty}^{t}[\hat{a}^{\da}(\tau)-\hat{a}(\tau)][\hat{I}(\tau),\hat{I}(t)] d\tau.
\end{align}
The last term in this expression is second order in $\lambda$. We thus average it with respect to the quantum conductor Hamiltonian isolated from the cavity, and use the expansion Eq.~\eqref{cur-corr-2} to rewrite
\begin{equation}\label{cur-cur}
\begin{split}
\langle [\hat{I}& (\tau),\hat{I}(t)] \rangle = \sum_{n \in \mathbb{Z}} \int \frac{d \omega}{2 \pi} S_n (\omega) \\ & \times \left(  e^{i \omega (t-\tau) - 2 i n \omega_0 t} - e^{-i \omega (t-\tau) - 2 i n \omega_0 \tau} \right)
\end{split}
\end{equation}
Our approach assumes small $\kappa_0$ and $\lambda$ such that the cavity field is essentially oscillating at the frequency $\omega_0$. After the change of variable 
\begin{equation} \label{chan-var}
\hat{a} (t) \rightarrow \hat{a} (t)e^{-i\omega_{0} t},
\end{equation}
$\hat{a} (t)$ becomes a slow field which evolves over time scales much larger than the decay of $[\hat{I}(\tau),\hat{I}(t)]$. We therefore replace $\hat{a}^{(\dagger)} (\tau)$ by $\hat{a}^{(\dagger)} (t)$ in the last term of Eq.~\eqref{eom-cav-1} which becomes
\begin{equation}\label{last}
\begin{split}
\dot{\hat{a}}(t) &= - \frac{\kappa_{0}}{2} \hat{a}(t) -\sqrt{\kappa_{0}}\, \hat{b}_{\rm in} (t) e^{i\omega_{0} t} + \lambda  \hat{I}(t)e^{i\omega_{0} t} \\
& + \lambda^2  C_{+} (t) \hat{a}^{\da} (t) - \lambda^2  C_{-} (t) \hat{a}(t),
\end{split}
\end{equation}
with 
\begin{equation}\label{c-inter}
C_{\pm} (t) = \int_{-\infty}^{t} d \tau e^{i\omega_{0} (t\pm \tau)} \langle [ \hat{I}(t), \hat{I}(\tau)] \rangle.
\end{equation}
Computing  $C_{\pm}$ with Eq.~\eqref{cur-cur}, we find an imaginary principal part term, describing frequency pull of the cavity, which we absorb into the cavity resonant frequency. We apply the rotating-wave approximation to the real part discarding all fast oscillating terms (beating with frequencies  multiple of  $\omega_0$) and obtain 
\begin{subequations}
\begin{align}
{\rm Re} \, C_+ & \simeq \frac{1}{2} \big[  S_1 (\omega_0) - S_1(\omega_0) \big ] = 0 \\
{\rm Re} \, C_- & \simeq \frac{1}{2} \big[ S_0 (\omega_0) - S_0(-\omega_0) \big]. 
\end{align}
\end{subequations}
We use these results in Eq.~\eqref{last} and go back to the original time frame for the cavity field $a(t) \rightarrow a(t)e^{i\omega_{0} t}$ to find the Heisenberg-Langevin equation
\begin{align}  \label{eom-cav-5}
\dot{\hat{a}}(t) &= -i\omega_{0}\hat{a}(t) - \frac{\kappa+\kappa_0}{2} \hat{a}(t) -\sqrt{\kappa_{0}} \, \hat{b}_{\rm in}(t) + \lambda \hat{I}(t),
\end{align}
advertised as Eq.~\eqref{langevin-mod} in the main text, where we introduced the cavity damping rate $\kappa = 2 \lambda^2 {\rm Re} \, C_-$. In the absence of external electromagnetic modes $\kappa_0=0$, it reduces to Eq.~\eqref{langevin} in the main text where the cavity  only interacts with the quantum conductor.

The solution to this equation can be obtained in time or frequency space. Combined with the boundary condition Eq.~\eqref{out-cav}, it leads to an expression for the output field in terms of the input field and the current in the quantum conductor
\begin{equation}
\hat{b}_{\rm out}(\omega) = \frac{\left[\omega-\omega_{0}+i(\kappa-\kappa_0)/2 \right]\hat{b}_{\rm in}(\omega) + i  \lambda \sqrt{\kappa_0} \, \hat{I} (\omega) }{\omega-\omega_{0}+i(\kappa+\kappa_0)/2}.
\end{equation}

\section{Finite-frequency Noise of a Quantum dot under ac excitation}

We consider a single-level quantum dot attached to two leads, denoted upper and lower leads, with tunnel couplings $t_U$ and $t_L$. Ignoring weak Coulomb interaction on the dot and assuming spinless electrons, the Hamiltonian takes the form
\begin{equation}
\begin{split}
H_{\rm dot} = \sum_{\alpha=U/L} \int d \varepsilon \, \varepsilon \, \hat{c}^\dagger_{\alpha} (\varepsilon) c_{\alpha}(\varepsilon)
+ \varepsilon_d d^\dagger d  \\
+ \sqrt{\nu_0} \sum_{\alpha=U/L} \int d \varepsilon \left( t_\alpha \hat{c}^\dagger_{\alpha} (\varepsilon) d + h.c. \right),
\end{split}
\end{equation}
with the constant density of states $\nu_0$ for the two leads. $\varepsilon_d$ is the dot single-level energy. $\hat{c}^\dagger_{\alpha}$ annihilates an electron with energy $\varepsilon$ in lead $\alpha=U/L$ and satisfies the anticommutation relation~\cite{blanter2000}
\begin{equation}
\{ \hat{c}^\dagger_\alpha (\varepsilon) \hat{c}_{\alpha'} (\varepsilon') \} = \delta_{\alpha,\alpha'} \delta ( \varepsilon - \varepsilon').
\end{equation}
The transport of electrons in this system is described by using scattering states. The scattering matrix is given by~\cite{Rothstein2009}
\begin{equation}\label{scattering}
  \begin{pmatrix} S_{UU} (\varepsilon) &  S_{UL} (\varepsilon) \\ S_{LU} (\varepsilon) &  S_{LL} (\varepsilon) 
  \end{pmatrix} = -1 + i g(\varepsilon)  \begin{pmatrix} \Gamma_U &  \sqrt{\Gamma_U \Gamma_L} \\ \sqrt{\Gamma_U \Gamma_L} & \Gamma_L 
  \end{pmatrix}
\end{equation}
with the function $g(\varepsilon) = (\varepsilon - \varepsilon_d + i \Gamma/2)^{-1}$. We have introduced the dot escape rates towards the two leads $\Gamma_\alpha = \pi \nu_0 |t_\alpha|^2$, and the total escape rate $\Gamma = \Gamma_L + \Gamma_U$. We write the current operator in the presence of an ac driving of the two leads following Ref.~\cite{pedersen1998}, where the scattering problem in the dot is disentangled from the potential modulation and the photo-excitation in the leads~\cite{lesovik1994}. Operators $\hat{c}'_{\alpha}(\varepsilon) $ are introduced, which describe scattering modes just after entering or leaving the dot. They are assumed to be adiabatically connected to the lead scattering modes through
\begin{equation}\label{adia}
\hat{c}'_{\alpha}(\varepsilon)  = \sum_{m \in \mathbb{Z}} \hat{c}_{\alpha}(\varepsilon-2  m \omega_0) c_m^\alpha
\end{equation}
where we introduced the Fourier coefficients for lead $\alpha$
\begin{equation}
e^{i e \int^t d t' \, V_{AC,\alpha} (t')} = \sum_{m \in \mathbb{Z}} c_m^\alpha e^{2 i m  \omega_0 t}.
\end{equation}
Eq.~\eqref{adia} describes the fact that an electron with energy $\varepsilon$ arriving at the dot-lead contact is composed by a coherent sum of lead electrons with energies $\varepsilon-2  m \omega_0$ which have absorbed or emitted $m$ photons from the ac field with probability amplitudes $c_m^\alpha$. Writting the current in terms of operators $\hat{c}'_{\alpha}(\varepsilon)$, as if no ac modulation were present, and substituting with Eq.~\eqref{adia}, one obtains the general expression for the current operator in lead $\alpha$ 
\begin{equation}\label{current2}
\begin{split}
\hat{I}_\alpha (t) = \frac{e}{h} \int \int d \varepsilon_1 d \varepsilon_2 \sum_{n_1,n_2} e^{i (\varepsilon_1-\varepsilon_2) t} \sum_{\gamma,\gamma'} c_{n_1}^\gamma c_{n_2}^{\gamma' *} \\ \times \hat{c}_{\gamma}^\dagger (\varepsilon_1-2 n_1 \omega_0) A_{\gamma \gamma'} (\alpha,\varepsilon_1,\varepsilon_2) \hat{c}_{\gamma'} (\varepsilon_2-2 n_2 \omega_0),
\end{split}
\end{equation}
where we use the scattering amplitudes
\begin{equation}
 A_{\gamma \gamma'} (\alpha,\varepsilon_1,\varepsilon_2) = \delta_{\gamma \gamma'} \delta_{\alpha \gamma'} - S^*_{\alpha \gamma} (\varepsilon_1)  S_{\alpha \gamma'} (\varepsilon_2)
\end{equation}
in terms of Eq.~\eqref{scattering}. Finally, the lead scattering operators have thermal distributions 
\begin{equation}
\langle \hat{c}^\dagger_\alpha (\varepsilon) \hat{c}_{\alpha'} (\varepsilon') \rangle = \delta_{\alpha,\alpha'} f_\alpha (\varepsilon) \delta ( \varepsilon - \varepsilon'),
\end{equation}
where the dc voltage $V_\alpha$ sets the chemical potential in lead $\alpha$, namely $f_\alpha(\varepsilon) = (1+e^{(\varepsilon-V_\alpha)/k_B T})^{-1}$. Note that we can always move the dc voltages $V_\alpha$ into the time dependence of Eq.~\eqref{current2}, instead of having them in the lead thermal distributions.

\subsection{Asymmetric quantum dot}

Expressions simplify in the limit of an asymmetric dot, $\Gamma_U \ll \Gamma_L$, more strongly coupled to the lower lead than to the upper lead. Computing the following scattering probabilities
\begin{subequations}
\begin{align}
|A_{L L} (U,\varepsilon_1,\varepsilon_2)|^2 & = \Gamma_U^2 \, \Gamma^2 \, |g(\varepsilon_1)|^2 |g(\varepsilon_2)|^2 \\ 
\begin{split}
|A_{U U} (U,\varepsilon_1,\varepsilon_2)|^2 & = |A_{L L} (U,\varepsilon_1,\varepsilon_2)|^2 \\&  + 4 \frac{\Gamma_U^2}{\Gamma^2} \sin^2 [\delta(\varepsilon_1) - \delta(\varepsilon_2) ],
\end{split}
\end{align}
\end{subequations}
where we defined the scattering phase $$\delta (\varepsilon) = \arctan [ \Gamma/(2 (\varepsilon_d - \varepsilon))],$$
we see that they both scale as $(\Gamma_U/\Gamma)^2$. In contrast to that, for $\Gamma_U \ll \Gamma_L$, we find
\begin{subequations}
\begin{align}
A_{U L} (U,\varepsilon_1,\varepsilon_2) \simeq i g(\varepsilon_2)\sqrt{\Gamma_U \Gamma}, \\
A_{L U} (U,\varepsilon_1,\varepsilon_2) \simeq -i g^*(\varepsilon_1)\sqrt{\Gamma_U \Gamma},
\end{align}
\end{subequations}
and their square modulus scale as $\Gamma_U/\Gamma$. We therefore neglect the $LL$ and $UU$ terms in the current. In addition, we assume that the voltage drop occurs at the upper dot-lead contact so that $c_m^L = \delta_{m,0}$ and we note $c_m^U \equiv c_m$. The dc voltage is $V \equiv V_U$ at the upper lead and $V_L=0$ in the lower lead. The potential of the dot is also to the ground. With these different notations and approximations, the current $\hat{I}_U (t)$ is given by Eq.~\eqref{currentasym} in the main text (see Methods) where the dc voltage is included in the operator time-dependence.

\subsection{Fully symmetric quantum dot}

We briefly discuss the case of a quantum dot symmetrically coupled to the leads $\Gamma_U = \Gamma_L$ and to the cavity.  The first order cavity-dot then takes the form $i \lambda (\hat{I}_U-\hat{I}_L) (\hat{a}^\dagger - \hat{a})$ involving the upper and lower currents with equal weights. The general approach devised for the tunnel junction also applies in this case, with the current $\hat{I}$ being replaced by $\hat{I}_U-\hat{I}_L$ in the Heisenberg-Langevin equation (Eq.~\eqref{langevin} in the main text). Setting 
\begin{equation}
\hat{I} = \hat{I}_U-\hat{I}_L
\end{equation}
and computing the correlators $\langle \hat{I} (\omega_1) \hat{I} (\omega_2) \rangle$, we obtain 
\begin{equation}
\begin{split}
& S_0 (\pm \omega_0) = \frac{e^2}{2 \pi \hbar^2}  \int d \varepsilon \sum_{n_1 n_2 n_3 \atop \gamma_1 \gamma_2} 
c_{n_1}^{\gamma_1} c_{n_2}^{\gamma_2 *} c_{n_3}^{\gamma_1 *}  c_{n_3+n_2-n_1}^{\gamma_2}  \\
& \times \bar{A}^*_{\gamma_1 \gamma_2} [\varepsilon+2 (n_3-n_1)\omega_0,\varepsilon \pm \omega_0+2 (n_3-n_1)\omega_0] \\[2mm] & \times \bar{A}_{\gamma_1 \gamma_2} (\varepsilon,\varepsilon \pm \omega_0) f_{\gamma_1}(\varepsilon-2 n_1 \omega_0) \\[1mm] & \times \bar{f}_{\gamma_2} (\varepsilon \pm \omega_0-2 n_2 \omega_0),
\end{split}
\end{equation}
with $\bar{f} (\varepsilon) = 1-f (\varepsilon)$ for the hole distribution, and
\begin{equation}
\begin{split}
& S_1 (\omega_0)  = \frac{e^2}{2 \pi \hbar^2}  \int d \varepsilon \sum_{n_1 n_2 n_3 \atop \gamma_1 \gamma_2} 
c_{n_1}^{\gamma_1} c_{n_2}^{\gamma_2 *} c_{n_3}^{\gamma_2}  c_{n_1+n_3-n_2+1}^{\gamma_1 *}  \\
& \times \bar{A}_{\gamma_2 \gamma_1} [\varepsilon+\omega_0+2 (n_3-n_2)\omega_0,\varepsilon +2 (n_3-n_2+1)\omega_0] \\[2mm] & \times \bar{A}_{\gamma_1 \gamma_2} (\varepsilon,\varepsilon + \omega_0) f_{\gamma_1}(\varepsilon-2 n_1 \omega_0) \\[1mm] & \times \bar{f}_{\gamma_2} (\varepsilon + \omega_0-2 n_2 \omega_0).
\end{split}
\end{equation}
We also introduced the scattering amplitudes 
\begin{equation}
\bar{A}_{\gamma_1 \gamma_2} (\varepsilon_1,\varepsilon_2) = A_{\gamma_1 \gamma_2} (U,\varepsilon_1,\varepsilon_2)-A_{\gamma_1 \gamma_2} (L,\varepsilon_1,\varepsilon_2).
\end{equation}
The knowledge of these correlators is sufficient to compute the quadrature variances of the cavity field, namely 
\begin{equation}\label{squeezed-quadra}
\Delta X_{1/2}^2 = \frac{S_0 (\omega_0) + S_0 (-\omega_0) \mp 2 {\rm Re}  \left[ S_1 (\omega_0)\right]}{S_0 (\omega_0) - S_0 (-\omega_0)}.
\end{equation}
We now turn to the numerical minimization of the variance $\Delta X_1^2$. The voltage is chosen to be zero on the dot so that the voltage potentials in the leads are opposite $V_U (t) = - V_L(t)$ by symmetry (the voltage drop is the same at each lead-dot contact). The applied voltage has a constant dc part and a sinusoidal part,  $V_U (t) = V + V_1 \cos ( 2 \omega_0 t)$, such that the $c_n$ coefficients are given by Bessel functions
\begin{equation}
c_n^{\gamma=U/L} = J_n \left( \pm \frac{ e V_1}{2 \hbar \omega_0} \right)
\end{equation}
At a general level, squeezing is optimized by setting $T=0$ and by having a sharp dot resonance $\Gamma/\omega_0 \ll 1$. In this fully symmetric situation however, the condition $e V = \hbar \omega_0$ is no longer met to optimize squeezing. Focusing on the asymptotic limit $\Gamma/\omega_0 \to 0$, we find the best squeezing
\begin{equation}
\Delta X_1^2 = 0.654 \qquad \textrm{for} \qquad \frac{e V_1}{2 \hbar \omega_0} =-0.58, 
\end{equation}
$\varepsilon_d =0$ and $e V = 0.6 \omega_0$. Squeezing is in fact not very sensitive to $V$ and $\varepsilon_d$ for $\Gamma \ll \omega_0$ as long as $|V| > |\varepsilon_d|$.
 

%

\end{document}